\documentclass[useAMS,usenatbib]{mn2e}

\usepackage{epsfig,color,pifont}
\usepackage{amsmath}

\usepackage{natbib}
\bibpunct{(}{)}{;}{a}{}{,}
\def\spose#1{\hbox to 0pt{#1\hss}}
\def\ltsimm{\mathrel{\spose{\lower 3pt\hbox{$\sim$}}
        \raise 2.0pt\hbox{$<$}}}
\def\gtsimm{\mathrel{\spose{\lower 3pt\hbox{$\sim$}}
        \raise 2.0pt\hbox{$>$}}}

\def\cm{{\rm\thinspace cm}}

\def\s{{\rm\thinspace s}}

\def\g{{\rm\thinspace g}}

\def\erg{{\rm\thinspace erg}}
\def\Hz{{\rm\thinspace Hz}}

\def\ster{{\rm\thinspace ster}}
\def\ergps{\hbox{${\rm\erg\s^{-1}\,}$}}

\def\pcm{\hbox{${\rm\cm^{-1}\,}$}}
\def\pcm2{\hbox{${\rm\cm^{-2}\,}$}}
\def\pcm3{\hbox{${\rm\cm^{-3}\,}$}}
\def\ergpscm3Hz{\hbox{${\rm\ergps\cm^{-3}\Hz^{-1}\,}$}}
\def\ergpscm3Hzster{\hbox{${\rm\ergps\cm^{-3}\Hz^{-1}\ster^{-1}\,}$}}
\def\gpcm3{\hbox{${\rm\g\cm^{-3}\,}$}}
\def\ergpcm2{\hbox{${\rm\erg\cm^{-2}\,}$}}
\def\ergpcm3{\hbox{${\rm\erg\cm^{-3}\,}$}}
\def\phpscm2{\hbox{${\rm photons\s^{-1}\cm^{-2}\,}$}}

\def\aap{{\rm A\&A}}
\def\apj{{\rm ApJ}}
\def\apjl{{\rm ApJL}}
\def\apjs{{\rm ApJS}}

\def\mnras{{\rm MNRAS}}

\def\araa{{\rm ARA\&A}}

\def\cpc{{\rm Comp.~Phys.~Comm.}}

\title[Adaptive image ray tracing]{Adaptive image ray-tracing for
  astrophysical simulations}

\author[E.~R.~Parkin]
       {E.~R.~Parkin\thanks{E-mail:parkin@astro.ulg.ac.be}\\ Institut
         d'Astrophysique et de G\'{e}ophysique, Universit\'{e} de
         Li\`{e}ge, 17, All\'{e}e du 6 Ao\^{u}t, B5c, B-4000 Sart
         Tilman, Belgium}

\begin{document}

\date{Accepted 2010 October 20. Received 2010 October 20; in original form 2010 July 20}

\pagerange{\pageref{firstpage}--\pageref{lastpage}} \pubyear{2010}

\maketitle

\label{firstpage}

\begin{abstract}
A technique is presented for producing synthetic images from numerical
simulations whereby the image resolution is adapted around prominent
features. In so doing, adaptive image ray-tracing (AIR) improves the
efficiency of a calculation by focusing computational effort where it
is needed most. The results of test calculations show that a factor of
$\gtsimm 4$ speed-up, and a commensurate reduction in the number of
pixels required in the final image, can be achieved compared to an
equivalent calculation with a fixed resolution image.
\end{abstract}


\section{Introduction}
\label{sec:intro}
Numerical models, such as hydrodynamical simulations, have become a
popular tool in astrophysical research. A common approach to
extracting observable quantities from a numerical simulation is to
trace the path of a ray through the simulation domain and sum-up the
value of some quantity at discrete intervals. Typically this is done
to integrate the column density through the simulation domain or to
solve the equation of radiative transfer to determine the emergent
intensity of emission. Synthetic images produced via ray-tracing can
be very useful for constraining model parameters, and examples of this
can be found in a wide range of astrophysical studies, e.g. symbiotic
recurrent nova \citep[][]{Orlando:2009, Drake:2010}, active galactic
nuclei \citep[][]{Bruggen:2009, Falceta:2010}, star formation
\citep[][]{Kurosawa:2004, Krumholz:2007, Parkin:2009b, Offner:2009,
  Peters:2010, Douglas:2010}, jets \citep[][]{Sutherland:2007,
  Bonito:2007, Saxton:2010}, starburst galactic winds
\citep{Cooper:2008}, and colliding winds binaries \citep{Pittard:2006,
  Pittard_Parkin:2010}. However, one drawback with using a fixed
resolution image (i.e. same pixel size at all points) is that a large
amount of computational effort can be expended on essentially blank
regions as features of interest rarely fill an entire image.

For example, in modern hydrodynamic simulations the grid/particle
resolution is often adapted to features in the flow. In grid-based
codes this is achieved using adaptive-mesh refinement \citep[AMR -
  e.g.][]{Berger:1989}, whereas smoothed-particle hydrodynamics
\citep[SPH - see ][]{Monaghan:1992} is inherently adaptive. When
producing an image the resolution must be sufficiently high to ensure
that the smallest scales of the simulation are well sampled. Yet there
may be regions of a simulation which do not warrant such a high level
of sampling. Furthermore, a ray which passes through a highly refined
region of the simulation domain may not necessarily hold much useful
information once it exits the grid. For instance, regions of high
intrinsic emission may be heavily absorbed such that the emergent
intensity is negligible. With these details in mind, as well as the
fact that as the computational requirements of simulations rise there
will be an associated increase in the memory required by datasets,
more efficient approaches to extracting observable quantities are
warranted.

This letter describes a method for adapting the resolution of a
ray-traced image to the feature(s) of interest. In so doing the
efficiency of a calculation is significantly improved. Adaptive image
ray-tracing (AIR) takes advantage of the varying simulation resolution
and magnitude of the {\it extracted} information encountered by a
ray. In essence, the advantages that adaptive mesh refinement (AMR)
has brought to grid-based hydrodynamical simulations are incorporated
into ray-tracing. The process is in a sense the reverse of
super-sampling \citep[e.g.][]{Whitted:1980, Genetti:1993}, where
instead of averaging over multiple rays, a single ray is
subdivided. Test calculations show that, compared to a fixed
resolution image, AIR provides an appreciable speed-up and a reduced
number of pixels for the resulting image. The remainder of this letter
is structured as follows: in \S~\ref{sec:air} the AIR technique is
outlined, \S~\ref{sec:results} presents results from test
calculations, and conclusions are presented in
\S~\ref{sec:conclusions}.

\section{Adaptive image ray-tracing}
\label{sec:air}

The basic principle behind producing a ray-traced image is to first
discretize the plane of the sky into a uniform array of pixels and
then follow the path of a ray for each respective pixel. The AIR
technique builds on this by taking an initially low resolution image
and then adapting (increasing) the resolution around sufficiently
prominent features of interest. This leads to a far more efficient
calculation as computational effort is concentrated on producing an
image of a desired variable(s). The structure of the AIR scheme is as
follows:
\begin{enumerate}
\item {\bf Construct base image:} After reading in the simulation file
  the base image resolution can be determined and the initial image
  constructed. As a guideline, the base image resolution can be set to
  sample the {\it lowest} resolution regions of the simulation
  domain. For example, for an AMR simulation setting the base image
  resolution equivalent to the base grid of the simulation is
  adequate.
\item {\bf Ray-trace:} Extract the desired information from the
  simulation domain by following the path of rays for each respective
  pixel. \label{step:raytrace}
\item {\bf Scan the image to check if refinement is required:} This is
  a relatively straightforward process of calculating the truncation
  error for a given pixel. The step in resolution between adjacent
  image pixels is prevented from being more than one refinement level
  - this ensures that the edges of features are well sampled.
\item {\bf Refine pixels:} The pixels selected by the truncation error
  check should now be refined. The $ij$ indexing of pixels cannot be
  preserved and therefore a heirarchical tree structure is
  required. Hence, during this step the book-keeping must be performed
  and information about the neighbours, parent, and children of a
  pixel updated.\label{step:refine}
\item {\bf Loop:} Repeat steps \ref{step:raytrace}-\ref{step:refine}
  until features in the image have been captured to the desired
  resolution and no more pixel refinement is required. For instance,
  the image resolution need not exceed that of the simulation.
\item {\bf Integrate quantities and output:} Once the final image has
  been constructed integrated quantities can be determined. An example
  of this would be broadband images in a given frequency/energy
  range. The sharpness of edges in the image can be improved by
  incorporating a super-resolution algorithm to interpolate between
  adjacent pixels of differing resolution
  \citep[e.g.][]{Chu:2009}. Following this, the final step is to
  output the image.
\end{enumerate}

As a note of caution, when calculating an integrated spectrum from an
image it is necessary to store the spectrum for each individual pixel
until the calculation is finished. This can lead to a prohibitive
overhead in memory requirements. However, this problem can be
alleviated by writing the spectrum for each pixel to temporary files
during the calculation.

Considering that the majority of the computational effort is expended
on the ray-tracing step, effective speed-up can be achieved by
parallelizing the AIR scheme on this step alone. This would involve
distributing the list of pixels across processors during each
ray-tracing sweep and then gathering the information for the
refinement check. Alternatively, it is conceivable that such a code
could be manufactured by taking an existing fixed-image ray-tracing
code and interfacing it with a parallel AMR library, e.g. {\sc
  PARAMESH} \citep{MacNeice:2000}, {\sc CHOMBO} \citep{Colella:2009},
{\sc DAGH} \citep{Parashar:1995}, {\sc SAMRAI}
\citep{Wissink:2001}. For instance, AMR libraries typically have
in-built functionality for parallel grid management, e.g.. refinement,
cell list book-keeping, interprocessor communication, and load
balancing. Therefore, ray-tracing could be handled on a pixel-by-pixel
basis using an existing fixed-image ray-tracing code, and the parallel
computation and grid management could be dealt with by functions
available in the AMR library. The process would be akin to the initial
refinement sweep performed in a grid-based hydrodynamics code, with
the difference that instead of populating a refined cell using the
simulation initial conditions a refined pixel is populated using the
results from a ray-tracing calculation.

\section{An example application}
\label{sec:results}

To demonstrate the advantages of using AIR compared to ray-tracing
with a fixed resolution image, calculations have been performed for a
test case. For this purpose a ring of hot gas residing at the centre
of a three-dimensional box has been simulated. The ring is aligned
with the $yz$-plane, has a constant thickness $w=1\times10^{17}\;{\rm
  cm}$, and has inner and outer radii of $r_{\rm
  i}=3.5\times10^{17}\;{\rm cm}$ and $r_{\rm o}=4\times10^{17}\;{\rm
  cm}$, respectively. The box has dimensions
$x=y=z=\pm5\times10^{17}\;{\rm cm}$. The gas density (g~cm$^{-3}$),
\begin{equation}
  \rho = \left\{ \begin{array}{lll} \beta r_{yz}^{-2} & ; r_{\rm
      i} \leq r_{yz} \leq r_{\rm o}, & |x| <
 w/2   \\ 1\times10^{-25}\;{\rm g~cm^{-3}} & ; {\rm otherwise} \\
  \end{array} \right. \label{eqn:rho}
\end{equation}
\noindent where $\beta=2.5\times10^{-9}\;{\rm g~cm^{-1}}$ and
$r_{yz}=\sqrt{y^2 + z^2}$. The gas temperature is $10^{8}\;$K and
$10^{4}\;$K inside and outside of the ring, respectively.

The ring and the box are modelled on an AMR grid constructed using the
{\sc FLASH} code v3.1.1 \citep{Fryxell:2000, Dubey:2009}, which
operates with the {\sc PARAMESH} block-structured AMR package
\citep{MacNeice:2000}. The coarse grid consists of $4^{3}$ blocks
containing $8^3$ cells. Refinement is performed on density and
temperature using 4 nested grid levels such that the effective
resolution is $512^{3}\;$cells.

For the AIR calculations the base image has a resolution of $32^2$
pixels and 4 nested levels of image refinement were used, giving an
effective image resolution of $512^2$ pixels (equivalent to a fixed
image calculation). The refinement check is performed on the
integrated 1-10~keV intrinsic X-ray flux using a modified
second-derivative interpolation error estimate
\citep{Lohner:1987}. This is essentially a second-order central
difference normalized by the sum of first-order forward and rearward
differences, which in one dimension on a uniform mesh is\footnote{In
  \citeauthor{Lohner:1987}'s formulation of the error estimator there
  is an additional term in the denominator added as a filter to
  prevent the refinement of ripples in hydrodynamic simulations. This
  term is not necessary for our purposes and, therefore, is not
  included.},
\begin{equation}
\xi_{i} = \frac{| p_{i+1} - 2p_{i} + p_{i-1}|}{|p_{i+1} - p_{i}| +
  |p_{i} - p_{i-1}|} \label{eqn:air_cond1}
\end{equation}
\noindent where $\xi$ is the truncation error, $p$ is the value of the
parameter on which image refinement is desired in pixel $i$
(e.g. X-ray flux). The multidimensional generalization of
Eq.~\ref{eqn:air_cond1} is found by taking all cross derivatives,
which leads to,
\begin{equation}
  \xi_{ij} = \left(\frac{\sum \limits_{uv}^{} \left(\frac{\partial^{2}p}{\partial
      x_u\partial x_v} \Delta x_u \Delta x_v\right)^2}{\sum\limits_{uv}^{}
    \left[\left(|\frac{\partial p}{\partial x_u}|_{i_u + 1/2} +
      |\frac{\partial p}{\partial x_u}|_{i_u - 1/2} \right)\Delta
      x_{u} \right]^2}\right)^{\frac{1}{2}} \label{eqn:air_cond2}
\end{equation}
\noindent where $u$ and $v$ are the image coordinates with indices $i$
and $j$, respectively, and $\Delta x_a$ is the separation of nodes in
the coordinate direction $a$. The partial derivatives are determined
at pixel centres and the sums are carried out over coordinate
directions. If $\xi_{ij} \ge \xi_{\rm crit}$ then the pixel is marked
for refinement. Determining the optimal value for $\xi_{\rm crit}$ for
a given application requires some experimentation; for the test case
values of $\xi_{\rm crit} > 0.01$ are effective at refining the ring
whilst maintaining efficiency. However, further tests performed on
images which contain more structure and varying degrees of contrast
reveal that $\xi_{\rm crit} \simeq 0.5$ is a more appropriate start
point. \citeauthor{Lohner:1987}'s error estimate is useful because it
is bounded (i.e. $0 \le \xi < 1$) so that a preset refinement
tolerance can be employed. It is also dimensionless, meaning that more
than one image parameter can be used to check for refiment without
encountering dimensioning problems.

To calculate the intrinsic X-ray emission we assume solar abundances
and use emissivities for optically thin gas in collisional ionization
equilibrium obtained from look-up tables calculated from the
\textsc{MEKAL} plasma code \citep{Kaastra:1992, Mewe:1995}.

\begin{figure}
  \begin{center}
    \begin{tabular}{c}
      \resizebox{60mm}{!}{\includegraphics{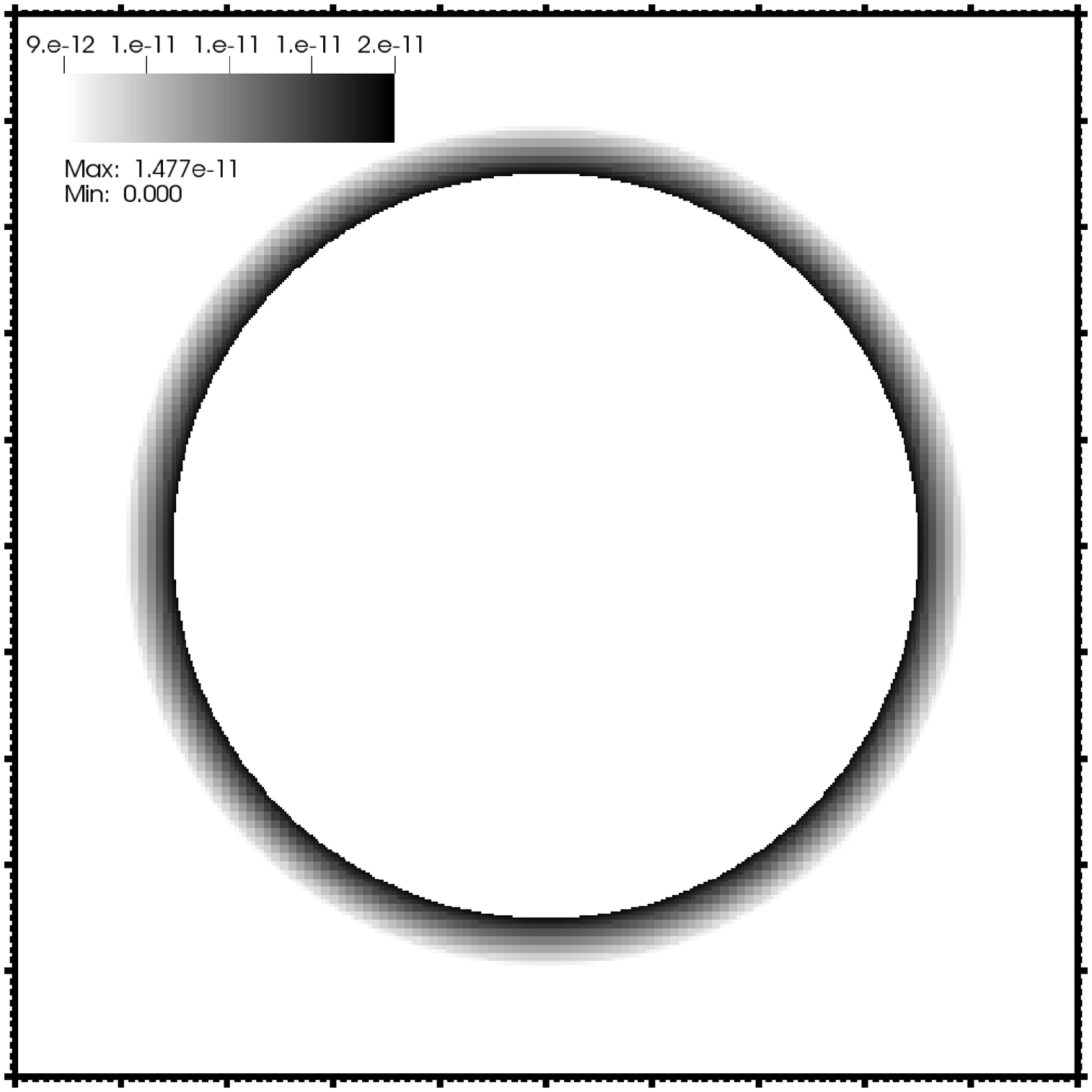}} \\
      \resizebox{60mm}{!}{\includegraphics{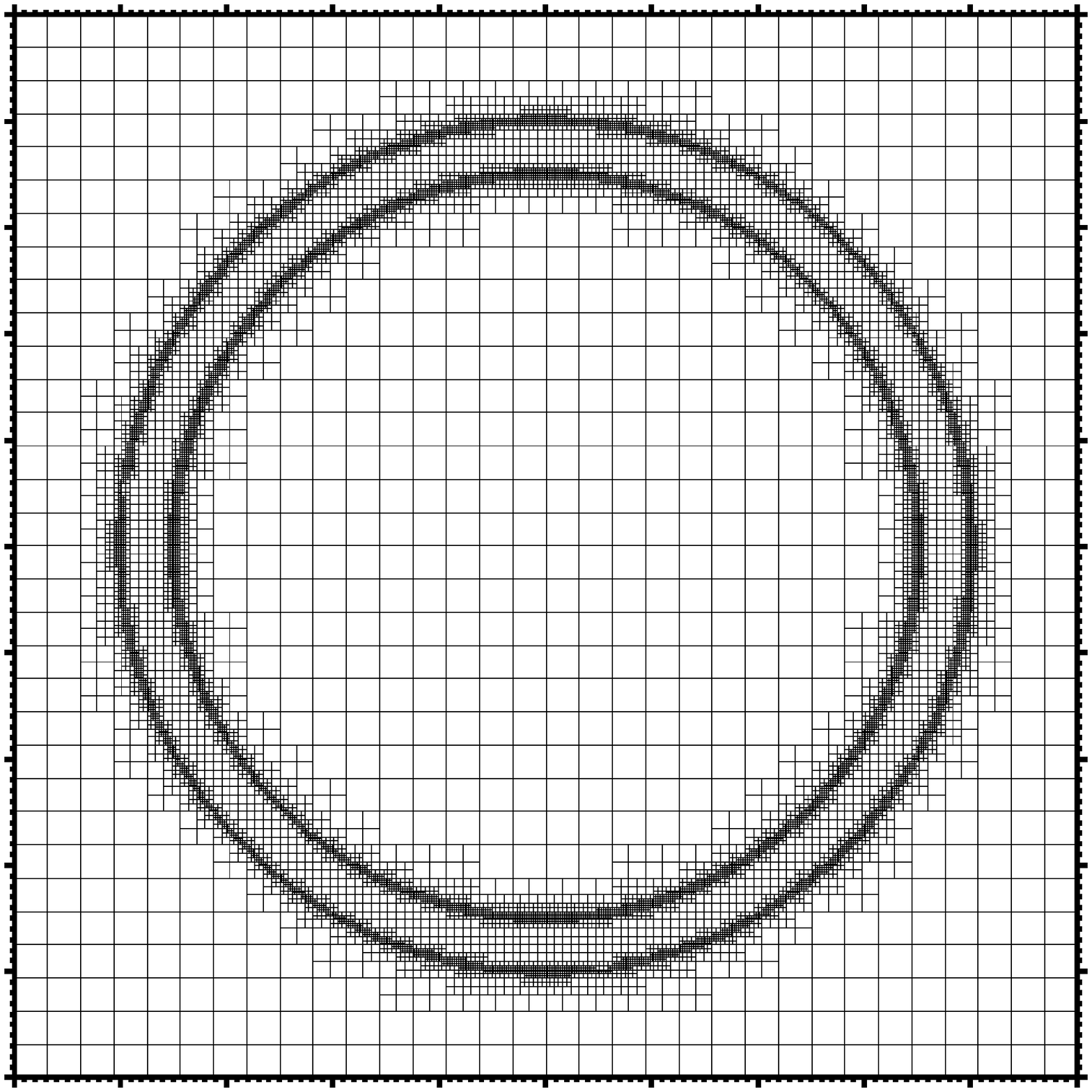}} \\
    \end{tabular}
    \caption{Results from the $\xi_{\rm crit}=0.8\;$ test calculation
      showing intrinsic 1-10 keV X-ray flux (top panel) and the image
      pixel mesh (lower panel). The images show a spatial extent of
      $u=v=\pm 5\times10^{17}\;{\rm cm}$ - large tick marks correspond
      to $1\times10^{17}\;{\rm cm}$. Note that for the adopted viewing
      angles $(u,v)=(y,z)$. For further details of the test
      calculations see \S~\ref{sec:results} and
      Table~\ref{tab:air_test}.}
    \label{fig:images}
  \end{center}
\end{figure}

The test problem places the observer viewing the ring face-on
(parallel to the $x$-axis) at a distance of 1~kpc and traces rays
through the hydrodynamic grid to calculate the intrinsic 1-10 keV
flux. For comparison, calculations have been performed with a fixed
image and with AIR using $\xi_{\rm crit}=0.2$, 0.5, and 0.8
(Fig.~\ref{fig:images} and Table~\ref{tab:air_test}). The fixed image
calculation is constrained by the fact that image pixels must be small
enough to sample the highest refined regions of the hydrodynamic grid,
hence the resolution must be $512\times512\;$pixels. In contrast, the
AIR calculation begins with a base image of $32\times32\;$pixels then
locates the feature of interest (the ring in this case) and increases
the image resolution respectively. This leads to far fewer pixels
being used and thus a more efficient calculation. For example, the
$\xi_{\rm crit}=0.8$ AIR calculation takes $\sim 1/7$ the time and
requires $\sim 1/12$ the image pixels, with a negligible error in the
integrated flux of 0.01\%.

As a approximate rule of thumb, AIR will reduce the calculation time
and pixel consumption by a factor of roughly the inverse of the image
filling factor, e.g. in the test calculation the ring fills $\sim 1/8$
of the image.

\begin{table}
\begin{center}
\caption[]{Adaptive image ray-tracing test calculation
  results. $\xi_{\rm crit}$ is the critical error estimate used to
  flag pixels for refinement (see Eq~\ref{eqn:air_cond2}), $F_{\rm X}$
  is the intrinsic 1-10 keV flux, $t$ is the time taken for the
  calculation. The error is calculated as $(F_{\rm X exact} - F_{\rm X
    ray})/F_{\rm X exact}$, where $F_{\rm X
    exact}=1.399586\times10^{-19}\;{\rm erg~s^{-1}~cm^{-2}}$ is the
  intrinsic flux summed from the hydrodynamic grid and $F_{\rm X ray}$
  is the ray-traced value for a given
  calculation.} \label{tab:air_test}
\begin{tabular}{lllll}
\hline
Calculation & $\xi_{\rm crit}$ & Fractional Error & $t$ & Pixels \\ 
            &                 &  & (s) & \\
\hline
Fixed & $-$ & $\sim0.$        & 9604 & 262144 \\
AIR   & 0.2 & $2\times10^{-6}$ & 2627 & 40512  \\
AIR   & 0.5 & $1\times10^{-4}$ & 1374 & 23384  \\
AIR   & 0.8 & $1\times10^{-4}$ & 1339 & 22748  \\
\hline
\end{tabular}
\end{center}
\end{table}

\section{Conclusions}
\label{sec:conclusions}

A method has been presented for adaptively increasing the resolution
of a ray-traced image around prominent features of interest. An
initially low resolution image is scanned and relevant pixels are
refined to increase the image resolution. The results of test
calculations show that considerable speed-up (a factor of $\sim 4-7$),
and a commensurate reduction in the number of pixels required for the
final image (a factor of $\sim 6-11$), can be achieved compared to an
equivalent calculation with a fixed resolution image. In conclusion,
adaptive image ray-tracing (AIR) improves the efficiency of a
calculation by focusing computational effort on extracting desired
information.

\subsection*{Acknowledgements}
This work was supported by a PRODEX XMM/Integral contract
(Belspo). The author thanks Julian Pittard and Andrii Elyiv for
insightful discussions and the anonymous referee for helpful comments
which improved the presentation of this paper. This research has used
software which was in part developed by the DOE-supported ASC/Alliance
Center for Astrophysical Thermonuclear Flashes at the University of
Chicago.


\label{lastpage}


\end{document}